\documentclass[preprint,showpacs,preprintnumbers,amsmath,amssymb]{revtex4}
\usepackage{graphicx}
\usepackage{dcolumn}
\usepackage{bm}

\begin{document}

\title{Instanton Solution of a Nonlinear Potential in
Finite Size}
\author{ Marco Zoli }
\affiliation{Istituto Nazionale Fisica della Materia -
Dipartimento di Fisica
\\ Universit\'a di Camerino, 62032, Italy. - marco.zoli@unicam.it}

\date{\today}

\begin{abstract}
The Euclidean path integral method is applied to a quantum
tunneling model which accounts for finite size ($L$)
effects. The general solution of the Euler Lagrange equation for
the double well potential is found in terms of Jacobi elliptic
functions. The antiperiodic instanton interpolates between the
potential minima at any finite $L$ inside the quantum regime and
generalizes the well known (anti)kink solution of the infinite
size case. The derivation of the
functional determinant, stemming from the quantum fluctuation contribution, is given in detail. The explicit formula for the finite size semiclassical path
integral is presented.
\end{abstract}

\maketitle

\section*{1. Introduction}

As a fundamental manifestation of quantum mechanics the tunneling
through a potential barrier plays a central role in many areas of
the physical sciences. Being intrinsically nonlinear phenomena,
quantum tunneling problems cannot be attacked by standard
perturbative techniques while semiclassical methods are known to
provide an adequate conceptual framework \cite{landau,berry}.

To be specific, I consider a particle of mass $M$ moving in a
bistable symmetric double well potential with oscillation
frequency $\omega$:

\begin{eqnarray}
V(x)=\,- {{M\omega^2} \over 2}x^2 + {{\delta} \over 4}x^4 +
{{\delta} \over 4}a^4 \label{eq:1}
\end{eqnarray}

The minima are located at $x=\, \pm a$  and the positive quartic
force constant $\delta$ (in units $eV \AA^{-4}$) is related to the
other potential parameters by $\delta=\, {M\omega^2/a^2}$. The
latter term in Eq.~(\ref{eq:1}) ensures that the potential is
positive defined and $V(x=\,\pm a)=\,0$. As it will be shown in
detail below, the minima are asymptotically connected by the
classical {\it (anti)kink} solution, a charge conserving domain
wall whose energy is inversely proportional to $\delta$ for a
given vibrational frequency inside the well. Thus, even small
quartic nonlinearities induce large classical energies which, in
turn, enhance the effects of the quantum fluctuations in the
overall probability amplitude for a particle to move from one
minimum to the next. This explains why, on general grounds,
perturbative methods fail to describe the tunneling physics.

In general, $\phi^4$ models also host potentials having a quartic
$\delta < 0$ \cite{i2} and positive sign of the quadratic term in
Eq.~(\ref{eq:1}). In this case the mid-point of the potential
valley would be a classically and locally stable configuration, a
metastable state from which the particle would tend to escape via
quantum tunneling. In such state, known in many branches of
physics as "false vacuum" \cite{coleman}, the system has a finite
lifetime due to the presence of a negative eigenvalue in the
quantum fluctuation spectrum which governs the decay into the
"true vacuum". The classical solution of the Euler-Lagrange
equation for metastable potentials is  a time-reversal invariant
{\it bounce} carrying zero charge. Also this object, whose
physical properties generally differ from the {\it (anti)kink},
has an energy inversely proportional to the anharmonic parameter
$\delta$ which makes perturbative analysis of metastability
unfeasible.

It has been noted since long that the $\delta=\,0$ singularity in
the thermodynamic limit of the partition function for the
anharmonic potential is formally similar to the condensation point
in the droplet model for phase nucleations \cite{langer}. The
nature of such singularity has been widely studied in the past
also in connection with fundamental investigations of the
anharmonic oscillator and its energy levels
\cite{bender,simon,sanchez}.

Tunneling problems in quantum field theory have been treated by
semiclassical methods \cite{dashen} which extend the WKB
approximation \cite{garg} and  determine the spectrum of
quadratic quantum fluctuations around the classical background.
While such methods had been first applied to infinite size
systems, generalizations to finite sizes \cite{lusch} have been
later on developed to obtain correlation functions and spectra of
$\phi^4$ potentials \cite{vale03}. The Casimir effect in $\phi^4$ theories is another remarkable example in which the properties of the system depend on the size of the cavity \cite{plun,lang}.

Finite size effects have become
a focus of research also on classical mechanical systems
\cite{garcia,faris} in which spatio-temporal noise of classical
origin induces escape from a metastable state. Examples of current
technological interest are the switching rate in the magnetization
of micromagnets \cite{braun} and the finite lifetime of metallic
nanowires shaped by cylinders whose radius undergoes fluctuations
of thermal origin \cite{yanson}. Such systems exhibit a structure
of thermal activation regimes which crucially depend on the system
size $(L)$ and the boundary conditions. The
transition in the classical activation rate versus $L$ is formally
identical to the crossover from the classical activation to the
quantum tunneling regime once $L$ is made proportional to the
inverse temperature of the system.

In general, the finite size formalism offers the bridge to derive
the thermodynamics of the system in the spirit of the
thermodynamic Bethe Ansatz \cite{zamol}: this is accomplished
by a Wick rotation which transforms the real time $t$ into the
imaginary time $\tau$ of the Matsubara formalism, $t\,\rightarrow
-i\tau$, where $\tau$ varies in a periodic spatial box of finite
size $L=\,\hbar \beta$ with $\beta=\,(K_BT)^{-1}$, $K_B$ being the
Boltzman constant and $T$ the temperature. Accordingly the
infinite size system maps onto the zero temperature limit.

With these premises, the path integral method in the Euclidean
version \cite{feyn,fehi} emerges as a natural tool to deal with
finite size/temperature tunneling phenomena in semiclassical
approximation. The quantum probability amplitude is given in terms
of a functional integral over all histories of the system depicted
by those $\tau-$ dependent paths which fulfill (anti)periodic
boundary conditions over a size $L$. In the semiclassical
treatment, the functional integral reduces to Gaussian path
integrals over quadratic quantum fluctuations around the classical
solution of the Euler-Lagrange equation. The fluctuations
contribution to the amplitude can be evaluated by a number of
techniques in the context of the functional determinants theory
\cite{gelfand,forman,mckane} that
implements the Euclidean path integral method. While the latter is
well known for the infinite size system (both for bistable and
metastable $\phi^4$ models) \cite{kleinert}, much less known is
the mathematics and the associated physics for the finite size
case. A contribution to the research field comes from this paper
which analyses the finite size Euclidean path integral for the
bistable $\phi^4$ model proposing a new family of classical
backgrounds. I emphasize that the model is intrinsecally non dissipative as there is no coupling to the heat bath while the temperature, as stated above, provides a measure of the system size. The work is organized as follows. Section 2 contains
the solution of the Euler-Lagrange equation, the finite
time instanton which minimizes the Euclidean action. This
family of paths, expressed in terms of Jacobi elliptic functions,
generalizes the (anti)kink peculiar of the infinite time theory and
interpolates between the minima of the potential
in Eq.~(\ref{eq:1}). Section 3 presents the generalities of the
semiclassical path integral formalism posing the problem of the
evaluation of the quantum fluctuations functional determinant. The
latter is derived in terms of elliptic integrals and analysed in
Section 4. Some final remarks are made in Section 5.

\section*{2. Classical Background}

Consider the potential of Eq.~(\ref{eq:1}). The real time
formalism does not allow classical motion in a double well
potential but, switching to the imaginary time formalism
$t\rightarrow -i\tau$, the potential is reversed
upside down and the classical orbit $x_{cl}(\tau)$ is obtained by
solving the classical equation of motion:

\begin{eqnarray}
M\ddot{x}_{cl}(\tau)=\,V'(x_{cl}(\tau)) \label{eq:2}
\end{eqnarray}

where $V'$ means derivative with respect to $x_{cl}$. Integrating
Eq.~(\ref{eq:2}), one gets

\begin{eqnarray}
{M \over 2}\dot{x}_{cl}^2(\tau) - V(x_{cl}(\tau))=\, E
\label{eq:2a}
\end{eqnarray}

with integration constant $E$. From Eq.~(\ref{eq:2a}) one easily
derives the equation of motion in integral form:

\begin{eqnarray}
\tau - \tau_0 =\,\pm \sqrt{{M \over 2}}
\int_{x_{cl}(\tau_0)}^{x_{cl}(\tau)} {{dx} \over {\sqrt{E +
V(x)}}} \label{eq:3}
\end{eqnarray}

The zero energy configuration is consistent with a physical
picture in which the particle starts on the top of a hill at
$x_{cl}(\tau =\,- \infty)=\, -a$ with zero velocity, crosses the
valley and reaches the top of the adjacent hill at $x_{cl}(\tau
=\,+ \infty)=\, +a$ with zero velocity. $\tau_0$ represents the
time at which the bottom of the valley (of the reversed potential)
is crossed and, by virtue of the $\tau-$ translational invariance,
it can be placed everywhere along the imaginary time axis. This
freedom has a price to be paid, that is a zero mode whose
eigenvalue causes a divergence in the path integral as discussed
below.

Imposing $x_{cl}(\tau_0)=\, 0$, one gets from Eq.~(\ref{eq:3})
with $E=\,0$, the well known (anti)kink

\begin{eqnarray}
x_{cl}(\tau)=\, \pm a \tanh[\omega (\tau - \tau_0)/\sqrt{2}]
\label{eq:4}
\end{eqnarray}

Since the transition happens in a short time, almost
instantaneously, the (anti)kink solutions are also named
(anti)instantons in quantum field theory
\cite{rajara,schaefer}. There are however other solutions of
Eq.~(\ref{eq:3}) associated to finite values of $E$ and
corresponding to the physical requirement that the classical path
has to connect the potential minima at finite times. Thus, the
finite size problem needs to be solved with antiperiodic boundary
conditions (APBC) such that $x_{cl}(\tau=\,\pm L/2)=\,\pm a$.
Define

\begin{eqnarray}
& &\chi_{cl}(\tau)=\, {1 \over {\sqrt{2}}}{{x_{cl}(\tau)} \over a}
\, \nonumber
\\
& &\kappa=\,{1 \over 2} + {{2E}\over {\delta a^4}}
\label{eq:5}
\end{eqnarray}

Then Eq.~(\ref{eq:3}) transforms into

\begin{eqnarray}
\tau - \tau_0 =\,\pm {{1 \over \omega}}
\int_{\chi_{cl}(\tau_0)}^{\chi_{cl}(\tau)} {{d\chi} \over
{\sqrt{\chi^4 - \chi^2 + \kappa/2 }}} \label{eq:6}
\end{eqnarray}

Solutions of Eq.~(\ref{eq:6}) can be found in three ranges of
$\kappa$:

\noindent{\bf 1) $\kappa \leq 0$}.  ${x_{cl}(\tau)}$ is unbounded
and always larger than $a$, hence it does not fulfill our physical
requirements.

\noindent{\bf 2) $0 < \kappa < 1/2$}. In this range and, for
$|\chi| \leq \sqrt{{1 - \sqrt{1 - 2\kappa}} \over 2}$, there is a
bounded, continuous and antiperiodic family of solutions given by

\begin{eqnarray}
& &x_{cl}(\tau)=\pm a F(\kappa) \cdot sn \Biggl({{\sqrt{{1 +
\sqrt{1 - 2\kappa}}}} \over {\sqrt{2}}} {{\omega}}(\tau - \tau_0),
p\Biggl)\, \nonumber
\\
& &p^2=\,{{\kappa} \over {\sqrt{{1 + \sqrt{1 - 2\kappa}}} -
\kappa}}; \,{} F(\kappa)=\,{{\sqrt{2\kappa}} \over {\sqrt{{1 +
\sqrt{1 - 2\kappa}}}}} \, \nonumber
\\
\label{eq:7}
\end{eqnarray}

This solution has often been proposed in the finite size studies
on the double well potential
\cite{liang,anker,maste}. Around this background one
can straightforwardly determine the low lying excitations in the
fluctuation spectrum by solving a standard Lam\'{e} equation
\cite{wang}. Note however that $F(\kappa)$ in Eq.~(\ref{eq:7}) is
always smaller than one. As the Jacobi elliptic sn-function is
$\leq 1$, the classical paths in Eq.~(\ref{eq:7})
undergo periodic oscillations between the slopes of the hills in
the reversed potential but never join the potential extrema
(unless the trivial case $\kappa=\, 1/2$ is recovered). This is
physically due to the fact that $\kappa < 1/2$ implies $E < 0$
(see Eq.~(\ref{eq:5})) but negative energies are not consistent
with finite path velocities on the $\tau-$ range boundaries. In
fact, from Eq.~(\ref{eq:2a}) it is easily seen that the relation

\begin{eqnarray}
{{\dot{x}_{cl}(\tau=\,\pm L/2)}}=\,\sqrt{2E/M} \label{eq:7a}
\end{eqnarray}

holds on the $\tau-$ boundaries. This observation alone proves
that, unlike it has been often assumed in the past, paths with $E
< 0$ cannot describe the physics of the  finite size double well
potential. Therefore Eq.~(\ref{eq:7}) is useless to our aim.

\noindent{\bf 3) $1/2 < \kappa$}. This turns out to be the
physically interesting case since in this range the classical
motion has $E > 0$. The solution of Eq.~(\ref{eq:6}) is presented
in detail.

The integral in the r.h.s. of Eq.~(\ref{eq:6}) yields:

\begin{eqnarray}
& &\int_{}^{} {{d\chi} \over {\sqrt{\chi^4 - \chi^2 + \kappa/2
}}}=\, {1\over 2} ^4\sqrt{2\over \kappa}F(\theta,s) \, \nonumber
\\
& &\theta=\,2\arctan\Bigl(^4 \sqrt{2\over \kappa} {}\chi \Bigr)\,
\nonumber
\\
& &s^2=\,{1\over 2}\biggl(1 + {1 \over {\sqrt{ 2\kappa}}}\biggr)
\label{eq:22}
\end{eqnarray}

where $F(\theta,s)$ is the elliptic integral of the first kind
with amplitude $\theta$ and modulus $s$:

\begin{eqnarray}
F(\theta,s)=\,\int_{0}^{\theta}{{d\alpha}\over {\sqrt{1 -
s^2\sin^2\alpha}}} \label{eq:23}
\end{eqnarray}

whose quarter-period is given by the first complete elliptic
integral $K(s)=\, F(\pi/2,s)$ \cite{wang,abram}. In terms of
$F(\theta,s)$, one defines the Jacobi elliptic functions
sn-amplitude, cn-amplitude, dn-amplitude :

\begin{eqnarray}
& & sn[F(\theta,s)]=\,\sin\theta\, \nonumber
\\
& & cn[F(\theta,s)]=\,\cos\theta\, \nonumber
\\
& & dn[F(\theta,s)]=\,\sqrt{1 - s^2 \sin^2\theta} \label{eq:24}
\end{eqnarray}

From Eqs.~(\ref{eq:6}),~(\ref{eq:22}) one gets:

\begin{eqnarray}
\pm  ^4\sqrt{\kappa \over 2 }2{{\omega}}(\tau - \tau_0)
=\,F(\theta,s) \label{eq:25}
\end{eqnarray}

and using the relations:

\begin{eqnarray}
& &\tan\theta=\,{{2\tan(\theta/2)} \over {1 - \tan^2(\theta/2)}}
\, \nonumber
\\
& &\tan\theta= {{sn[F(\theta,s)]} \over {cn[F(\theta,s)]}}
\label{eq:26}
\end{eqnarray}

Eq.~(\ref{eq:25}) transforms into:

\begin{eqnarray}
& &{{sn(2\varpi,s)} \over {cn(2\varpi,s)}}= \,{{2 \cdot \,{}
^4\sqrt{2\over \kappa} {}\chi_{cl}} \over {1 - \sqrt{2\over
\kappa} {}\chi_{cl}^2}}\, \nonumber
\\
& &\varpi=\,\pm ^4\sqrt{\kappa \over 2 }{{\omega}}(\tau - \tau_0)
\label{eq:27}
\end{eqnarray}

The condition $^4 \sqrt{2\over \kappa} {}\chi_{cl}(\tau) \neq \pm
1$ imposed by Eq.~(\ref{eq:27}) is equivalent to $x_{cl}(\tau)/a
\neq \pm (2s^2-1)^{-1/2}$ which is always fulfilled in the
physical range of our interest. The l.h.s. in the first of
Eq.~(\ref{eq:27}) can be rewritten using the double arguments
relations:

\begin{eqnarray}
& &sn(2\varpi,s)=\,{{2sn(\varpi,s)cn(\varpi,s)dn(\varpi,s)} \over
{1 - s^2sn^4(\varpi,s)}} \, \nonumber
\\
& &cn(2\varpi,s)=\,{{cn^2(\varpi,s) -
sn^2(\varpi,s)dn^2(\varpi,s)} \over {1 - s^2sn^4(\varpi,s)}} \,
\nonumber
\\
\label{eq:28}
\end{eqnarray}

and, through the definition in Eq.~(\ref{eq:5}), from
Eqs.~(\ref{eq:27}),~(\ref{eq:28}) I derive the two solutions for
the classical equation of motion in the finite size model:

\begin{eqnarray}
& &x^{(1)}_{cl}(\tau)=\, ^4\sqrt{2\kappa}\,{} a
{{sn(\varpi,s)dn(\varpi,s)} \over {cn(\varpi,s)}}\, \nonumber
\\
& &x^{(2)}_{cl}(\tau)=\,- ^4\sqrt{2\kappa}\,{} a {{cn(\varpi,s)}
\over {sn(\varpi,s)dn(\varpi,s)}}\, \nonumber
\\
\label{eq:8}
\end{eqnarray}

\begin{figure}
\includegraphics[height=7cm,angle=-90]{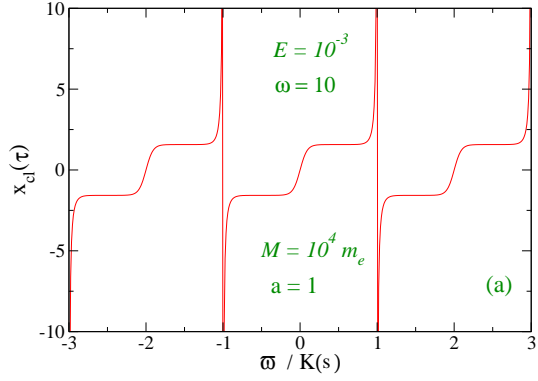}
\caption{\label{fig:1} (Color online) Classical path
$x^{(1)}_{cl}(\tau)$ in Eq.~(\ref{eq:8}), for a choice of energy
$E$ (in meV) and potential parameters: $a$ (in $\AA$),
$\hbar\omega$ (in meV) and particle mass $M$ in units of the
electron mass. The center of the (anti)instantons is taken at
$\tau_0=\,0$.}
\end{figure}

\begin{figure}
\includegraphics[height=7cm,angle=-90]{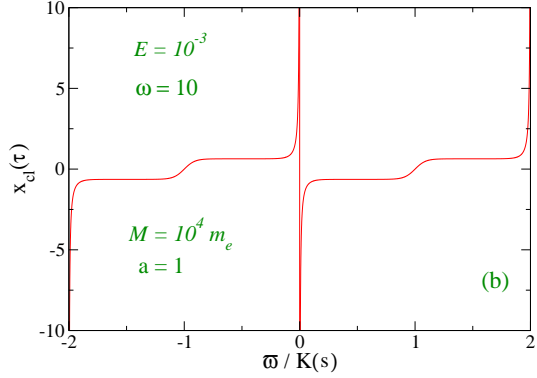}
\caption{\label{fig:2} (Color online) Classical path
$x^{(2)}_{cl}(\tau)$ in Eq.~(\ref{eq:8}), for the same parameters
as Fig.~\ref{fig:1}.}
\end{figure}

The antiperiodic paths of Eq.~(\ref{eq:8}) are plotted in
Fig.~\ref{fig:1} and Fig.~\ref{fig:2} respectively, for a choice
of potential parameters $\omega$, $M$, $a$ and energy $E$ which
fulfill the criteria for the validity of the semiclassical
approximation. Aside from the prefactor $^4\sqrt{2\kappa}\,{} a$,
the two families of classical paths are antireciprocal of each
other. Thus the zeros of $x^{(1)}_{cl}(\tau)$, occuring at
$\varpi=\, \pm 2nK(s)$ with integer $n$  would be divergencies for
$x^{(2)}_{cl}(\tau)$. Vice-versa, the zeros of
$x^{(2)}_{cl}(\tau)$ at $\varpi=\, \pm (2n + 1)K(s)$ are singular
points for $x^{(1)}_{cl}(\tau)$.

In fact such singularities, that remind us of the caustic points
in the classical orbit of the harmonic oscillator \cite{schulman},
happen to be outside the range physically relevant to our
purposes. I clarify this point.

First, note that the solutions in Eq.~(\ref{eq:8}) have a period
$4K(s)$ thus, in principle,  they should be discussed in the range
$\varpi \in [-2K(s), 2K(s)]$ since the amplitude $\theta$ of the
elliptic integral varies as $\theta \in [-\pi, \pi]$. Second,
focus on $x^{(1)}_{cl}(\tau)$ and apply the APBC,
$x^{(1)}_{cl}(\tau=\,\pm L/2)=\,\pm a$, emboding the physics of
the finite size problem. Setting $\tau_0 =\,0$ without loss of
generality, the first of Eq.~(\ref{eq:8}) yields:

\begin{eqnarray}
& &1=\, ^4\sqrt{2\kappa}\,{} {{sn(\varpi^*,s)dn(\varpi^*,s)} \over
{cn(\varpi^*,s)}} \, \nonumber
\\
& &\varpi^*=\,\pm ^4\sqrt{\kappa \over 2 }{{\omega}}{L \over 2}
\label{eq:9}
\end{eqnarray}

Representing the elliptic functions in Fourier series \cite{wang},
I calculate Eq.~(\ref{eq:8}) and  obtain the values $\varpi^*$
fulfilling the APBC. Precisely, the computational method sets a
value of $E$ and determines the corresponding $\varpi^*$ such as
Eq.~(\ref{eq:9}) holds. In this way the program calculates the
period $L=\,L(E)$ and establishes the one to one correspondence
between $E$ and $\omega^*$ at which the path interpolates
between the minima. The ratio $\varpi^*/K(s)$ turns out to be
always smaller than one, accordingly the singularities at
$\varpi=\,\pm K(s)$ are physically avoided. In other words, the
path $x^{(1)}_{cl}(\tau)$ always connects the potential minima
well before encountering the singular points. Thus, taking the
center of motion at $x_{cl}(\tau_0=\,0)=\,0$, $x^{(1)}_{cl}(\tau)$
properly describes a family of finite size (anti)instantons which
are bounded, continuous and odd in the range $\tau \in
[-L/2,L/2]$.

An analogous procedure with APBC applied to $x^{(2)}_{cl}(\tau)$
in Fig.~\ref{fig:2} leads to determine the boundaries $\pm
\varpi^*/K(s)$. As $\tau_0$ may be located everywhere along the
$\tau-$ axis one may define in principle an even range
$\varpi/K(s)$ with no singularities also for $x^{(2)}_{cl}(\tau)$.
However, setting $\tau_0=\,0$, the choice of $x^{(1)}_{cl}(\tau)$
happens to be more convenient since the latter is well behaved for
$\tau \in [-L/2,L/2]$.

The remarkable fact is that the $(\omega^*, E)$ correspondences
obtained by the two families of classical backgrounds are
essentially identical. This means that the two independent
solutions of the finite size classical equations of motion predict
the same physics.

Correctly, in the limit $E \rightarrow 0 \, \, \,(\kappa
\rightarrow 1/2)$, one gets from Eq.~(\ref{eq:8}):

\begin{eqnarray}
& &x^{(1)}_{cl}(\tau) \rightarrow \, \pm a \tanh[\omega (\tau -
\tau_0)/\sqrt{2}] \, \nonumber
\\
& &x^{(2)}_{cl}(\tau) \rightarrow \, \mp a \coth[\omega (\tau -
 \tau_0)/\sqrt{2}] \label{eq:10}
\end{eqnarray}

The first in Eq.~(\ref{eq:10}) is the (anti)kink whose center of
motion is set at $x^{(1)}_{cl}(\tau_0)=\,0$ whereas the second is
the solution of the infinite size classical equation
Eq.~(\ref{eq:2}) fulfilling the condition
$x^{(2)}_{cl}(\tau_0)=\,\infty$.

Further, using the relations:

\begin{eqnarray}
& &{{d\,{} sn(\varpi,s)}\over
{d\varpi}}=\,cn(\varpi,s)dn(\varpi,s) \, \nonumber
\\
& &{{d\,{} cn(\varpi,s)}\over
{d\varpi}}=\,-sn(\varpi,s)dn(\varpi,s) \, \nonumber
\\
& &{{d\,{} dn(\varpi,s)}\over {d\varpi}}=\,-s^2
sn(\varpi,s)cn(\varpi,s) \label{eq:30}
\end{eqnarray}

one gets, from the first in Eq.~(\ref{eq:8}), the particle path
velocity as:

\begin{eqnarray}
{{ \dot{x}_{cl}(\tau)}}=\, a \omega \sqrt{\kappa} \Biggl[1 - 2 s^2
sn^2(\varpi,s) +  {{sn^2(\varpi,s)dn^2(\varpi,s)} \over
{cn^2(\varpi,s)}} \Biggr]\, \nonumber
\\
\label{eq:31}
\end{eqnarray}

Using the second of Eq.~(\ref{eq:27}), one can easily check that
Eq.~(\ref{eq:31}) does not vanish on the $\tau$-range boundaries
consistently with the general constraint expressed by
Eq.~(\ref{eq:7a}). The latter, together with Eq.~(\ref{eq:31}),
permits to define the one to one correspondence between $E$ and
$\omega^*$.

\section*{3. Semiclassical Path Integral  }

In the semiclassical model, the particle path $x(\tau)$ splits in
the classical background $x_{cl}(\tau)$ plus the quantum
fluctuations $\eta(\tau)$. Thus, the path integral for space-time
particle propagation between the positions $x_i$ and $x_f$ in an
imaginary time $L$ is:

\begin{eqnarray}
<x_f | x_i>_L=\,\exp\biggl[- {{A[x_{cl}} \over {\hbar}} \biggr]
\cdot \int D\eta(\tau) \exp\biggl[- {{A_f[\eta]} \over {\hbar}}
 \biggr]\, \nonumber
\\
\label{eq:16}
\end{eqnarray}

The classical action $A[x_{cl}]$ is formally given by

\begin{eqnarray}
A[x_{cl}]=\,\sqrt{2M} \int_{-a}^{a}dx_{cl} \sqrt{E + V(x_{cl})} -
E \cdot L(E) \, \nonumber
\\
\label{eq:11}
\end{eqnarray}

Inserting the potential in Eq.~(\ref{eq:1}), Eq.~(\ref{eq:11})
transforms into:

\begin{eqnarray}
& &A[x_{cl}]=\,A_0 \int_0^1 dy \sqrt{{{4E} \over {\delta a^4}} +
(1 + y)^2(1 - y)^2} - E \cdot L(E) \, \nonumber
\\
& &A_0=\,{{\sqrt{2} \delta a^4} \over \omega} \label{eq:12}
\end{eqnarray}

suitable for computation while, in the limit $E \rightarrow 0$,
one obtains the well known result:

\begin{eqnarray}
A[x_{cl}]&\rightarrow&\, {{2 \sqrt{2}} \over 3} {{M^2 \omega^3}
\over \delta} \label{eq:13}
\end{eqnarray}

Eq.~(\ref{eq:13}) represents the (anti)kink energy (in units of
$\hbar$) whose inverse dependence on $\delta$ has been pointed out
in the Introduction.

Being quadratic in the quantum fluctuations, $A_f[\eta]$ defines a
stability equation whose eigenvalues represent the fluctuation
spectrum \cite{kleinert}. Carrying out Gaussian path integrals
over the fluctuation paths, one gets formally

\begin{eqnarray}
& &\int D\eta(\tau) \exp\biggl[- {1 \over {\hbar}} A_f[\eta]
\biggr]= \, \aleph \cdot Det\Bigl[\hat{O}\Bigr]^{-1/2} \,
\nonumber
\\
& &\hat{O}\equiv -\partial_{\tau}^2 + {{ V''(x_{cl}(\tau))}/ M}
\label{eq:17}
\end{eqnarray}

where $\aleph$ depends on the measure of integration. The heart of
the matter is the evaluation of the functional determinant in
Eq.~(\ref{eq:17}). Three observations permit to get the clue to
the problem.

\noindent{\bf A)} Differentiating Eq.~(\ref{eq:2}) with respect to
$\tau$, one notes that ${\dot{x}_{cl}(\tau)}$ solves the
homogeneous equation associated to the second order
Schr\"{o}dinger-like differential operator $\hat{O}$

\begin{eqnarray}
\hat{O} \eta(\tau)=\,0 \label{eq:18}
\end{eqnarray}

As a general consequence of the $\tau$-translational invariance, a
Goldstone mode appears in the system and this is true both for the
infinite and the finite size cases. Thus, the particle path
velocity is proportional to the fluctuation which makes
$A_f[\eta]$ vanishing, ${\dot{x}_{cl}(\tau)} \propto \eta_0(\tau)$
with eigenvalue $\varepsilon_0=\,0$.

\noindent{\bf B)} As ${{x}_{cl}(\tau)}$ is monotonic within a
period $L$, ${\dot{x}_{cl}(\tau)}$ has no nodes hence,
$\eta_0(\tau)$ represents the ground state and $\varepsilon_0$ is
the lowest eigenvalue in the fluctuation spectrum in agreement
with the fact that the potential is stable. However, the zero mode
eigenvalue breaks the Gaussian approximation and makes the path
integral divergent. The trouble is overcome by regularizing the
fluctuations determinant, extracting $\varepsilon_0$ from
Eq.~(\ref{eq:17}) and evaluating its contribution separately
resorting to collective coordinates \cite{langer,kleinert}. The
replacement yields, for the infinite and finite size theories
respectively:

\begin{eqnarray}
& &(\varepsilon_0)^{-1/2} \rightarrow \, \sqrt{{{A[x_{cl}]}\over
{2\pi\hbar}}}L  \,{}\,\,\,\,\, {infinite \,\,\, size}\, \nonumber
\\
& &(\varepsilon_0)^{-1/2} \rightarrow \, \sqrt{{{A[x_{cl}] +
E\cdot L }\over {2\pi\hbar}}}L \, {} \,\,\,\,\, {finite \,\,\,
size}\, \nonumber
\\
\label{eq:18a}
\end{eqnarray}

Note incidentally that also the {\it infinite size} instantonic
theory assumes, strictly speaking, that the size $L$ along the
$\tau-$ axis is {\it finite} (although large) so that
$\varepsilon_0^{-1/2} \propto L$ and the path integral is finite.
Otherwise the zero mode singularity could not be removed. This
observation points to a formal contradiction of the {\it infinite
size} theory that, in fact, stimulated me to develop a consistent
description in which both the classical background and the zero
mode of the quantum fluctuations can be precisely determined at
any finite size.

\noindent{\bf C)} As stated below Eq.~(\ref{eq:31}),
$\dot{x}_{cl}(\tau)$ takes the same values on the $\tau-$
boundaries. More generally, for any two points $\varpi_1,
\varpi_2$ such that $\varpi_2=\,\varpi_1 \pm 2K(s)$, from
Eq.~(\ref{eq:31}) one gets:
$\dot{x}_{cl}(\varpi_2)=\,\dot{x}_{cl}(\varpi_1)$. As the path
velocity {\it is} (proportional to) the ground state fluctuation,
the important consequence is that the fluctuation eigenmodes obey
periodic boundary conditions (PBC).

Collecting the informations from {\bf A)} to {\bf C)} and using a
fundamental result of the theory of functional determinants with
PBC \cite{forman}, I rewrite the fluctuation determinant in
Eq.~(\ref{eq:17}) as

\begin{eqnarray}
& &Det[\hat{O}]=\,\varepsilon_0 \cdot Det^R[\hat{O}] \, \nonumber
\\
& &Det^R[\hat{O}]=\,{{<\eta_0 |\eta_0> \bigl(\eta_1(\varpi_2) -
\eta_1(\varpi_1)\bigr)} \over {{\eta_0(\varpi_1) W(\eta_0,
\eta_1)}}} \label{eq:19}
\end{eqnarray}

Where $Det^R[\hat{O}]$ indicates that the determinant is
regularized \cite{mckane} after extracting the zero
eigenvalue. $\eta_0$ and $\eta_1$ are independent solutions of
Eq.~(\ref{eq:18}):

\begin{eqnarray}
& &\eta_0=\,\dot{x}_{cl}\, \nonumber
\\
& &\eta_1=\,{{\partial {x}_{cl}} \over {\partial m}}; \, {} \,
m=\,s^2 \label{eq:20}
\end{eqnarray}

$W(\eta_0, \eta_1)$ is their Wronskian and $<\eta_0 |\eta_0>$ is
the squared norm. Using the first in Eq.~(\ref{eq:8}) and
Eq.~(\ref{eq:20}), $Det^R[\hat{O}]$ can be worked out
analytically.

\section*{4. Regularized Fluctuation Determinant }

I show now in detail how to obtain the regularized
$Det^R[\hat{O}]$ of Eq.~(\ref{eq:19}) on the base of the Wronskian
construction. The two points $\varpi_2, \varpi_1$ map onto $L/2$
and $-L/2$ along the $\tau-$ axis according to the definition in
Eq.~(\ref{eq:27}).

{\bf I)} The $\tau-$ derivatives of the two independent solutions
$\eta_0$ and $\eta_1$ of Eq.~(\ref{eq:20}) read respectively:

\begin{eqnarray}
& &\dot{\eta}_0(\tau)=\,{\ddot{x}_{cl}(\tau)} \, \nonumber
\\
& &\dot{\eta}_1(\tau)=\,{\partial \over {\partial m
}}\Bigl[{{\partial {x}_{cl}(\tau)} \over {\partial \varpi}} \cdot
{{\partial \varpi} \over {\partial \tau}}\Bigr] \label{eq:32}
\end{eqnarray}

Then, the Wronskian $W(\eta_0, \eta_1)$ is:

\begin{eqnarray}
W(\eta_0, \eta_1)=\,\dot{x}_{cl}(\tau) {{\partial
\dot{x}_{cl}(\tau)}\over {\partial m}} - {\ddot{x}_{cl}(\tau)}
{{\partial {x}_{cl}(\tau)}\over {\partial m}}\, \nonumber
\\
\label{eq:33}
\end{eqnarray}

The Wronskian is constant and therefore can be calculated in any
convenient $\tau \in [-L/2,L/2]$. The center of the classical
paths, located at $\tau_0$, is the best choice as
$\ddot{x}_{cl}(\tau_0) \equiv \, {{x}_{cl}(\tau_0)=\,0}$ and
$\dot{x}_{cl}(\tau_0)=\, a\omega \sqrt{\kappa}$.

Observing that:

\begin{eqnarray}
& &{{\partial \varpi} \over {\partial \tau}}=\,{{\omega} \over {2
\sqrt{s^2 - 1/2}}}\, \nonumber
\\
& &{\partial \over {\partial m}}\Bigl[{{\partial \varpi} \over
{\partial \tau}}\Bigr]=\,- {\omega \over 4}(s^2 - 1/2)^{-3/2}\,
\nonumber
\\
\label{eq:34}
\end{eqnarray}

from Eqs.~(\ref{eq:32}) and ~(\ref{eq:33}), I get

\begin{eqnarray} W(\eta_0, \eta_1)=\,-
{{\omega^2 a^2} \over {16}}(s^2 - 1/2)^{-3} \label{eq:35}
\end{eqnarray}

{\bf II)} I evaluate $\eta_1(\tau=\,\pm L/2)$ writing $\eta_1$ as

\begin{eqnarray}
& &\eta_1(\tau)=\,{{\partial {x}_c(\tau)}\over {\partial \varpi}
}\cdot {{\partial \varpi} \over {\partial m }} \, \nonumber
\\
& &\varpi=\,{1 \over 2}F(\theta,s) \label{eq:36}
\end{eqnarray}

Using the relations:

\begin{eqnarray}
& &{{\partial \varpi}\over {\partial m}}=\,{1 \over
{4s}}{{\partial F(\theta,s)}\over {\partial s}}\, \nonumber
\\
& &{{\partial F(\theta,s)}\over {\partial s}} =\,{s \over
{\bar{s}^2}}\Biggl[ {{E(\theta,s) - \bar{s}^2 F(\theta,s)} \over
{s^2}} -  {{\sin\theta \cos\theta} \over {\sqrt{1 -
s^2\sin^2\theta}}} \Biggr]\, \nonumber
\\
& &E(\theta,s)=\,\int_{0}^{\theta}{{d\alpha}{\sqrt{1 -
s^2\sin^2\alpha}}} \, \nonumber
\\
& &\bar{s}^2=\,1 - s^2 \label{eq:37}
\end{eqnarray}

I derive the following boundary properties:

\begin{eqnarray}
& &\theta(L/2)=\, 2 arctan\Biggl({1 \over { \,{}^4 \sqrt{2
\kappa}}} \Biggr)=\, - \theta(-L/2)\, \nonumber
\\
& &F\bigl(\theta(L/2),s\bigr)=\,-F\bigl(\theta(-L/2),s\bigr)\,
\nonumber
\\
& &E\bigl(\theta(L/2),s\bigr)=\,-E\bigl(\theta(-L/2),s\bigr)\,
\nonumber
\\
\label{eq:38}\end{eqnarray}

After setting $\tau_0=\,0$, it follows that:

\begin{eqnarray}
& &\eta_1(L/2)=\,-\eta_1(-L/2)\, \nonumber
\\
& &\eta_1(-L/2)=\,\dot{x}_c(-L/2) \Bigl[\biggl({{\partial
\varpi}\over {\partial \tau} }\biggr)^{-1} \cdot {{\partial
\varpi} \over {\partial m }}\Bigr]_{\tau=\,-L/2} \, \nonumber
\\
& &\dot{x}_c(\pm L/2)=\,2a \omega \sqrt{\kappa}\,s^2 \bigl(1 -
sn^2(\varpi^*)\bigr)\, \nonumber
\\
& &sn^2(\varpi^*)=\,{{1 - \cos[\theta(L/2)]} \over {1 + \sqrt{1 -
s^2 \sin^2[\theta(L/2)]}}}\, \nonumber
\\
& &\varpi^*=\, \pm {{\omega L} \over {4 \sqrt{s^2 -
1/2}}}
\label{eq:39}
\end{eqnarray}

The last in Eq.~(\ref{eq:39}) coincides with $\varpi^*$ in
Eq.~(\ref{eq:9}). Note that $\eta_1$ is a solution of
Eq.~(\ref{eq:18}) but, unlike $\eta_0$, it does not represent a
quantum fluctuation component. This explains why $\eta_1$ is odd
on the boundaries whereas $\eta_0$ fulfills PBC. Thus,
$Det^R[\hat{O}]$ in Eq.~(\ref{eq:19}) can be rewritten as:

\begin{eqnarray}
Det^R[\hat{O}]=\, {2{{<\eta_0 |\eta_0>} \over  {W(\eta_0,
\eta_1)}}} \Bigl({{\partial \varpi} \over {\partial
\tau}}\Bigr)^{-1} {{\partial \varpi}\over {\partial
m}}\Bigl|_{\tau=\,L/2} \, \nonumber
\\
\label{eq:40}\end{eqnarray}

The last factor in Eq.~(\ref{eq:40}) is computed via
Eqs.~(\ref{eq:37}),~(\ref{eq:38}) and turns out to be positive
thus ensuring the correct (negative) sign to $Det^R[\hat{O}]$ (due
to $W(\eta_0, \eta_1)<\,0$). It is well known from the general
theory that only ratios of functional determinants have physical
meaning both in value and sign. In fact being a product of  an
infinite number of eigenvalues with modulus larger than one, each
determinant is separately divergent in the $L \rightarrow \infty$ limit.
To get a finite ratio, I normalize $Det^R[\hat{O}]$ over the
harmonic oscillator determinant $Det[\hat{h}]$ which, for periodic
boundary conditions, reads:

\begin{eqnarray}
& &Det[\hat{h}]=\,-4\sinh^2(\omega L/2) \, \nonumber
\\
& &\hat{h}\equiv \, -\partial^2_{\tau} + \omega^2 \label{eq:41}
\end{eqnarray}

As $Det[\hat{h}]$ and $Det^R[\hat{O}]$ have the same exponential
divergence, their ratio is finite in the $L \rightarrow \infty$ limit.
Moreover it is positive,  consistently with the fact that the
contribution by the zero mode eigenvalue  (extracted from
$Det[\hat{O}]$) is positive. Therefore also
$Det[\hat{O}]/Det[\hat{h}]$ has to be positive since the system is
stable and there are no lower eigenvalues below the zero mode
eigenvalue.

{\bf III)} $\eta_0$ is proportional to the ground state ortonormal
component in the series expansion for the path fluctuation. I
define $\bar{\eta}_0=\,\bar{N} \eta_0$ such that, in
Eq.~(\ref{eq:40}), $<\eta_0|\eta_0>\equiv\,\bar{N}^{-2}$. Then,
using Eq.~(\ref{eq:31}), I derive the squared norm in the form

\begin{eqnarray}
\bar{N}^{-2}&=&\,G \int_0^{\varpi^*}d\varpi \Biggl[1 - 2 s^2
sn^2(\varpi,s) + {{sn^2(\varpi,s)dn^2(\varpi,s)} \over
{cn^2(\varpi,s)}} \Biggr]^2\, \nonumber
\\
G&=&\,{{\omega a^2} \over {2 (s^2 - 1/2)^{3/2}}}
\label{eq:42}\end{eqnarray}

suitable for computation. Alternatively, one may proceed through
the definition of the classical action in Eq.~(\ref{eq:11}) and
get:

\begin{eqnarray}
\bar{N}^{-2}=\,{1 \over M} \Bigl({{A[x_{cl}]}} +
{E \cdot L(E)}\Bigr) \label{eq:43}
\end{eqnarray}

In any case $<\eta_0|\eta_0>$ has the dimension of $[\omega a^2]$
hence, from Eqs.~(\ref{eq:34}),~(\ref{eq:35}),~(\ref{eq:42}), it
follows that $Det^R[\hat{O}]$ is proportional to $\omega^{-2}$ and
correctly it does not carry any dependence on $a$. This completes
the analysis since all ingredients required to compute
Eq.~(\ref{eq:40}) are now known.

Taking the ratio between the determinants in Eq.~(\ref{eq:40}) and
Eq.~(\ref{eq:41}), in the $L \rightarrow \infty$ limit, I get

\begin{eqnarray}
{{Det^R[\hat{O}]} \over {Det[\hat{h}]}} \rightarrow {1 \over {12
\omega^2}} \label{eq:43a}
\end{eqnarray}

thus recovering the well known value  of the infinite size
instantonic approach. This proves the correctness of the
analytical procedure.

Then, the semiclassical path integral in Eq.~(\ref{eq:16}) for one
(anti)instanton takes the final expression:

\begin{eqnarray}
& &<x_f | x_i>_L=\,\sqrt{{M \over {2\pi \hbar L}}} \cdot {1 \over
{2 \sinh\bigl(\omega L/{2}\bigr)}} \Omega(L) L \, \nonumber
\\
& &\Omega(L)=\,\exp\biggl[- {{A[x_{cl}]} \over {\hbar}}  \biggr]
\cdot \Omega^{QF}(L)  \, \nonumber
\\
& &\Omega^{QF}(L)=\, \sqrt{{{M \bar{N}^{-2}}\over {2\pi\hbar}}}
\sqrt{{{Det[\hat{h}]} \over {Det^R [\hat{O}]}}} \label{eq:21}
\end{eqnarray}

which can be evaluated using
Eq.~(\ref{eq:12}),~(\ref{eq:40}),~(\ref{eq:41}),~(\ref{eq:43}).
Note that: {\it i)} The harmonic determinant is, in turn,
normalized over the free particle determinant
$Det[-\partial^2_{\tau}]=\, \sqrt{{M \over {2 \pi \hbar L}}} $
which incorporates the constant $\aleph$. {\it ii)}
$\Omega^{QF}(L)$ accounts for the quantum fluctuation effects in
the finite size tunneling problem. {\it iii)} $\Omega(L)$ is the
overall tunneling frequency which removes the twofold degeneracy
of the double well potential in a finite domain. From the path
integral in Eq.~(\ref{eq:21}) one can extract the physical
properties of the closed quantum system.

\section*{5. Final Remarks}

This work presents the mathematical description of the finite
size semiclassical theory for the bistable $\phi^4$
potential based on the Euclidean path integral formalism. In
particular the two key ingredients of the model, namely classical
equation of motion and quantum fluctuation spectrum, are analysed
in detail. I have studied the classical equation of motion and
obtained the family of classical paths which interpolates between
the potential minima in the finite size system: such paths are
necessarily associated to positive classical energies and fulfill
antiperiodic boundary conditions. Around such non trivial background one may derive the mass spectrum in $\phi^4$ theories \cite{toharia} by solving a stability equation more general than the standard Lam\`{e} equation associated to Eq.~(\ref{eq:7}) \cite{harrin}.

Due to the time translational invariance, the spectrum of the
quadratic quantum fluctuations contains a zero mode which breaks
the Gaussian approximation and makes the path integral divergent.
The result of the regularization procedure for the finite size
case is presented. Using the theory of the functional
determinants, I have derived the regularized fluctuations
determinant giving the full set of equation required to compute
it. Finally, I have evaluated the contribution to the path
integral originating from the quantum fluctuations and obtained an
explicit expression for the tunneling energy in the finite size
bistable $\phi^4$ model. This study permits to compute, for
specific choices of potential parameters, the finite size induced
renormalization of the tunneling energy with respect to the
predictions of the infinite size instantonic approach. In view of the mapping to the temperature axis allowed by the Matsubara formalism, the obtained results may be well applied to calculate, for instance, the thermodynamical properties of systems described by Ginzburg-Landau field theories \cite{krum,zinn}, the finite $T$ splitting energy in two level systems arising in amorphous compounds \cite{zawa,io3} and materials with local structural instabilities \cite{yuand,io4,io5} as well as in solids showing macroscopic quantum tunneling of magnetization \cite{stampchud}.

Along similar semiclassical patterns one can study a metastable $\phi^4$
model in finite size. For the latter one has to build the
generalized bounce solution (whose properties however differ very
much from those of the instanton) and analyse the quantum
fluctuation spectrum with particular care to the ground state
negative eigenvalue which causes metastability. The softening of
the low lying fluctuation modes together with the decay rate for a
finite size system will be the subject of a next work.


\begin{references}
\bibitem{landau}
L.D.Landau, E.M.Lifshitz, {\it Quantum Mechanics} 3rd Edition,
Butterworth-Heinemann, Oxford (1977).
\bibitem{berry}
M.V.Berry, K.E.Mount, Rep. Prog. Phys. {\bf 35},  315  (1972).
\bibitem{i2}
M.Zoli, Phys. Rev. B {\bf 72},  214302 (2005).
\bibitem{coleman}
S.Coleman, Phys.\ Rev.\ D {\bf 15},  2929  (1977); C.G.Callan,
S.Coleman, Phys.\ Rev.\ D {\bf 16},   1762 (1977).
\bibitem{langer}
J.S.Langer, Ann. Phys. {\bf 41},  108  (1967).
\bibitem{bender}
C.M.Bender and T.T.Wu,   Phys.\ Rev.\ Lett. {\bf 21},  406  (1968);
ibid. Phys.\ Rev. {\bf 184},  1231  (1969).
\bibitem{simon}
B.Simon, Ann. Phys. {\bf 58},  76  (1970).
\bibitem{sanchez}
A.M.Sanchez and J.D.Bejarano,  J.Phys.A: Math.Gen. {\bf 19},
887 (1986).
\bibitem{dashen}
R.Dashen, B.Hasslacher and A.Neveu, Phys.\ Rev.\ D {\bf 10},
4114 (1974); ibid. {\bf 11},  3424 (1975); ibid. {\bf 12},  2443  (1975).
\bibitem{garg}
A.Garg, Am.J.Phys. {\bf 68}, 430 (2000).
\bibitem{lusch}
M.L\"{u}scher,  Commun. Math. Phys. {\bf 104},  177  (1986).
\bibitem{vale03}
G.Mussardo, V.Riva and G.Sotkov, Nucl. Phys. B {\bf 670}, 464 (2003).
\bibitem{plun}
G.Plunien, B.M\"{u}ller, W.Greiner, Phys. Rep. {\bf 134}, 87 (1986).
\bibitem{lang}
K.Langfeld, F.Schm\"{u}ser, H.Reinhardt, Phys.\ Rev.\ D {\bf 51}, 765 (1995).
\bibitem{garcia}
J.Garcia-Ojalvo, J.M.Sancho, {\it Noise in Spatially Extended
Systems}, Springer, N.Y./Berlin (1999).
\bibitem{faris}
W.G.Faris and G.Jona-Lasinio, J.Phys.A {\bf 15},   3025 (1982).
\bibitem{braun}
H.-B.Braun,  Phys.\ Rev.\ Lett. {\bf 71},   3557 (1993).
\bibitem{yanson}
A.I.Yanson, I.K.Yanson and J.M. van Ruitenbeck, Nature {\bf 400},
144 (1999); ibid. Phys.\ Rev.\ Lett. {\bf 84},  5832  (2000).
\bibitem{zamol}
A.Zamolodchikov,  J.Phys.A:Math.Gen. {\bf 39},  12863 (2006).
\bibitem{feyn}
R.P.Feynman,  Rev.\ Mod.\ Phys. {\bf 20},  367 (1948).
\bibitem{fehi}
R.P.Feynman and A.R.Hibbs, {\it Quantum Mechanics and Path
Integrals} Mc Graw-Hill, New York (1965).
\bibitem{gelfand}
I.M.Gelfand and A.M.Yaglom, J.Math.Phys. {\bf 1},  48 (1960).
\bibitem{forman}
R.Forman, Invent.Math. {\bf 88}, 447 (1987).
\bibitem{mckane}
A.J.McKane and M.B.Tarlie, J.Math.Phys. {\bf 28}  (1995) 6931.
\bibitem{kleinert}
H.Kleinert, {\it Path Integrals in Quantum Mechanics, Statistics,
Polymer Physycs and Financial Markets}, World Scientific
Publishing, Singapore (2004).
\bibitem{rajara}
R.Rajaraman, {\it Solitons and Instantons} North Holland,
Amsterdam (1982).
\bibitem{schaefer}
T.Sch\"{a}fer and E.V.Shuryak, Rev.Mod.Phys. {\bf 70}, 323 (1998).
\bibitem{liang}
J.-Q.Liang, H.J.W.M\"{u}ller-Kirsten, D.K.Park and F.Zimmerschied
Phys.\ Rev.\ Lett. {\bf 81}, 216 (1998).
\bibitem{anker}
J.Ankerhold and H.Grabert, Phys.\ Rev.\ E {\bf 61}, 3450 (2000).
\bibitem{maste}
R.S.Maier and D.L.Stein,  Phys.\ Rev.\ Lett.  {\bf 87},
270601 (2001).
\bibitem{wang}
Z.X.Wang and D.R.Guo,  {\it Special Functions},  World
Scientific, Singapore (1989).
\bibitem{abram}
M.Abramowitz and I.A.Stegun,  {\it Handbook of Mathematical
Functions},   Dover Publications, New York (1972).
\bibitem{schulman}
L.S.Schulman, {\it Techniques and Applications of Path
Integration}, Wiley\&Sons, New York (1981).
\bibitem{toharia}
B.Grzadkowski, M.Toharia, CERN-PH-TH/2004-003
\bibitem{harrin}
B.J.Harrington, Phys.\ Rev.\ D {\bf 18}, 2982  (1978).
\bibitem{krum}
J.A.Krumhansl, J.R.Schrieffer, Phys.\ Rev.\ B {\bf 11}, 3535  (1975).
\bibitem{zinn}
J. Zinn-Justin, {\it Quantum Field Theory and Critical Phenomena} Clarendon Press, Oxford (1989).
\bibitem{zawa}
K.Vlad\'{a}r, A.Zawadowski,  Phys.\ Rev.\ B {\bf 28}, 1564 (1983);
ibid. {\bf 28}, 1582 (1983); ibid. {\bf 28}, 1596 (1983).
\bibitem{io3}
M.Zoli, Acta Physica Polonica A {\bf 77}, 639 (1990).
\bibitem{yuand}
C.C.Yu, P.W.Anderson, Phys.Rev.B {\bf 29}, 6165 (1984).
\bibitem{io4}
M.Zoli, Phys.Rev.B  {\bf 44}, R7163 (1991).
\bibitem{io5}
M.Zoli, in {\it Lattice Effects in High $T_c$ Superconductors},
       eds. Y.Bar-Yam, T.Egami, J.Mustre de Leon, A.R.Bishop.
       World Scientific, Singapore (1992) p.195.
\bibitem{stampchud}
P.C.E.Stamp, E.M.Chudnovsky, B.Barbara, Int. J. Mod. Phys.B {\bf 6}, 1355 (1992).

\end{references}
\end{document}